\documentclass[12pt]{article}

\begin{document}
\tolerance=5000
\def\be{\begin{equation}}
\def\ee{\end{equation}}
\def\bea{\begin{eqnarray}}
\def\eea{\end{eqnarray}}
\def\nn{\nonumber \\}
\def\cF{{\cal F}}
\def\det{{\rm det\,}}
\def\Tr{{\rm Tr\,}}
\def\e{{\rm e}}
\def\etal{{\it et al.}}
\def\erp2{{\rm e}^{2\rho}}
\def\erm2{{\rm e}^{-2\rho}}
\def\er4{{\rm e}^{4\rho}}
\def\etal{{\it et al.}}
\def\gsim{\ ^>\llap{$_\sim$}\ }

\ 

\vskip -3cm

\  \hfill
\begin{minipage}{2.5cm}
January 2002 \\
\end{minipage}

\vfill

\begin{center}
{\large\bf Asymptotically de Sitter dilatonic space-time, 
holographic RG flow and
conformal anomaly from (dilatonic) dS/CFT correspondence}

\vfill

{\sc Shin'ichi NOJIRI}\footnote{nojiri@cc.nda.ac.jp} 
and {\sc Sergei D. ODINTSOV}$^{\spadesuit}$\footnote{
odintsov@ifug5.ugto.mx, odintsov@mail.tomsknet.ru} \\

\vfill

{\sl Department of Applied Physics \\
National Defence Academy,
Hashirimizu Yokosuka 239-8686, JAPAN}

\vfill

{\sl $\spadesuit$
Tomsk State Pedagogical University, Tomsk, RUSSIA and
Instituto de Fisica de la Universidad de Guanajuato, \\
Lomas del Bosque 103, Apdo. Postal E-143, 
37150 Leon,Gto., MEXICO}

\vfill

{\bf ABSTRACT}

\end{center}

The number of asymptotically de Sitter (non-singular) solutions of 5d 
dilatonic gravity with positive cosmological constant is found.
These solutions are similar to the previously known asymptotically AdS 
spaces where dilaton may generate the singularity. Using these solutions
the consistent $c$-function is proposed in the same way as in AdS/CFT.
 The consistency of RG flow gives further support for dS/CFT correspondence.
>From holographic RG flow equations we calculate the holographic 4d 
conformal anomaly with dilatonic contributions. This conformal anomaly 
turns out to be the same as in AdS/CFT correspondence.

\vfill


\newpage

The universality of holographic principle manifests itself in the form
of AdS/CFT correspondence \cite{AdS} as well as recently proposed 
dS/CFT correspondence \cite{ds1,strominger}. Despite the numerous attempts
(for not complete list of refs. on dS/CFT correspondence, see 
\cite{ds2,ds3a,NOds}) dS/CFT set-up is mainly studied from the
gravitational side and it is not well understood. One of the fundamental 
problems in the development of dS/CFT is the lack of the suitable 
dual CFT which is presumably Euclidean and not unitary. 
Probably it is related with the fact that de Sitter space unlike to AdS 
is very difficult to realize as vacuum space of string theory.
Nevertheless, it is became clear that most of relations discussed 
in AdS/CFT have their analogs also in dS/CFT \cite{ds3a}.

In the present letter we extend the study of holographic RG flows 
developed in AdS/CFT to asymptotically de Sitter dilatonic spaces.
In particulary, the asymptotically de Sitter (non-singular) solution of 5d 
dilatonic gravity is found. Using this solution (with non-trivial dilaton)
the definition of c-function in dS/CFT correspondence is proposed.
The proposed c-function which is similar (up to the sign of the potential)
to the AdS c-function gives further support for dS/CFT set-up.
The holographic RG flow equations are written. Using them the dilatonic 
conformal anomaly which turns out to be the same as in AdS/CFT is found.

We start from the following action of dilatonic gravity
in $d+1$ dimensions:
\be
\label{i}
S=-{1 \over 16\pi G}\int d^{d+1}x \sqrt{-G}\left(R - \Lambda 
 - \alpha G^{\mu\nu}\partial_\mu \phi \partial_\nu \phi \right)\ .
\ee
We now assume $\lambda^2\equiv \Lambda$ to be positive. 

 From the variation of the action (\ref{i}) with respect to 
the metric $G^{\mu\nu}$, one obtains\footnote{
The conventions of curvatures are given by
\begin{eqnarray*}
R&=&G^{\mu\nu}R_{\mu\nu} \\
R_{\mu\nu}&=& -\Gamma^\lambda_{\mu\lambda,\kappa}
+ \Gamma^\lambda_{\mu\kappa,\lambda}
 - \Gamma^\eta_{\mu\lambda}\Gamma^\lambda_{\kappa\eta}
+ \Gamma^\eta_{\mu\kappa}\Gamma^\lambda_{\lambda\eta} \\
\Gamma^\eta_{\mu\lambda}&=&{1 \over 2}G^{\eta\nu}\left(
G_{\mu\nu,\lambda} + G_{\lambda\nu,\mu} - G_{\mu\lambda,\nu} 
\right)\ .
\end{eqnarray*}
}
\be
\label{iit}
0=R_{\mu\nu}-{1 \over 2}G_{\mu\nu}R + {\Lambda \over 2}G_{\mu\nu}
 - \alpha \left(\partial_\mu\phi\partial_\nu\phi 
 -{1 \over 2}G_{\mu\nu}G^{\rho\sigma}\partial_\rho \phi
\partial_\sigma \phi \right)
\ee
and from that of dilaton $\phi$
\be
\label{iiit}
0=\partial_\mu\left(\sqrt{-G}G^{\mu\nu}\partial_\nu\phi\right)\ .
\ee
We assume that $\phi$ depends 
only on the one of the coordinate, say $y\equiv x^d$ 
and we also assume that 
$G_{\mu\nu}$ has the following form
\be
\label{viit}
ds_{d+1}^2=\sum_{\mu,\nu=1}^{d+1} G_{\mu\nu}dx^\mu dx^\nu
=f(y)dy^2 + y\sum_{i,j=1}^d g_{ij}dx^i dx^j
\ee
Here $g_{ij}$ is the metric in the Einstein manifold, which 
is defined by
\be
\label{vat}
r_{ij}=kg_{ij}\ .
\ee
 $r_{ij}$ is the Ricci tensor given by $g_{ij}$ and $k$ is 
a constant, especially $k>0$ for sphere and $k=0$ for the 
flat Minkowski space and $k<0$ for hyperboloid. 

The equations of motion (\ref{iit}) and (\ref{iiit}) take the 
following forms:
\bea
\label{vit}
0&=&{1 \over 2}{rf \over y}
 -{d(d-1) \over 8}{1 \over y^2} - {\lambda^2 \over 2}f 
+ {\alpha \over 2}(\phi')^2 \\
\label{viitb}
0&=&-\left(r_{ij} -{1 \over 2}rg_{ij}\right){f \over y} \nn
&& + \left\{{d-1 \over 4}{f'\over fy}
 -{(d-1)(d-4) \over 8}{1 \over y^2} 
- {\lambda^2 \over 2}f - {\alpha \over 2}(\phi')^2\right\}g_{ij} \\
\label{viiit}
0&=&\left(\sqrt{{y^d \over f}}\phi'\right)'\ .
\eea
Here $'$ expresses the derivative with respect to $y$ and 
$r\equiv g^{ij}r_{ij}=kd$. 
Eq.(\ref{vit}) corresponds to $(\mu,\nu)=(d,d)$ in (\ref{iit}) and 
Eq.(\ref{viitb}) to $(\mu,\nu)=(i,j)$. The case of $(\mu,\nu)=(0,i)$ 
or $(i,0)$ is identically satisfied. Integrating (\ref{viiit}), 
one finds
\be
\label{ixt}
\phi'=c\sqrt{-{f \over y^d}}\ .
\ee
Here $c$ is some constant.
Substituting (\ref{ixt}) into (\ref{vit}), we can solve it 
algebraically with respect to $f$:
\be
\label{xt}
f=-{d(d-1)  \over 4y^2
\lambda^2 \left(1 - {\alpha c^2 \over \lambda^2 y^d}
 - {kd \over \lambda^2 y}\right)}\ .
\ee
Then we find from (\ref{ixt}) and (\ref{xt}), 
\be
\label{xiiit}
\phi=c\int dy \sqrt{{d(d-1) \over 
4y^{d+2}\lambda^2 \left(1 - {\alpha c^2 \over \lambda^2 y^d}
- {kd \over \lambda^2 y}\right)}}\ .
\ee

The EMT $T_{\mu\nu}^\phi$ coming from $\phi$ is, 
as one can find it from (\ref{iit}), given by 
$T_{\mu\nu}^\phi=\alpha \left(\partial_\mu\phi\partial_\nu\phi 
 -{1 \over 2}G_{\mu\nu}G^{\rho\sigma}\partial_\rho \phi
\partial_\sigma \phi \right)$. Using (\ref{ixt}) one gets 
\be
\label{EC}
T_{yy}=-{\alpha c^2 f \over 2 y^d}\ ,\quad 
T_{ij}={\alpha c^2 \over 2y^{d-1}}g_{ij}\ .
\ee
The pressure corresponding to $T_{ij}$ is always positive 
and the energy density corresponding to $T_{yy}$ is positive as 
long as $f$  (\ref{xt}) is negative. Then the strong and 
weak energy conditions are satisfied as long as $f$ is positive.

When $y$ is small, $f(y)$ (\ref{xt}) behaves as
\be
\label{SG1}
f(y)\sim {d(d-1) y^{d-2} \over 4\alpha c^2}\ ,
\ee
which makes a curvature singularity at $y=0$.
The scalar curvature behaves when $y\sim 0$ as
\be
\label{Axxxviib}
R \sim \alpha c^2 y^{-d} \ .
\ee
The curvature singularity can be generated by the singular 
behavior of the dilaton $\phi$ when $y\sim 0$:
\be
\label{SG2}
\phi(y) \sim {\rm sgn}(c)\sqrt{d(d-1) \over 4\alpha} \ln y\ .
\ee
Here ${\rm sgn}(c)$ expresses the sign of $c$:
\be
\label{SG3}
{\rm sgn}(c)=\left\{\begin{array}{lll}
+1\ &\ \mbox{if}\ &\ c>0 \\
 -1\ &\ \mbox{if}\ &\ c<0 \\
\end{array}\right.\ .
\ee
On the other hand, when $y\rightarrow +\infty$, 
by using Eq.(\ref{xt}), one finds that $f(y)$ behaves as
\be
\label{ff1}
f=-{d(d-1)  \over 4y^2
\lambda^2 }\ .
\ee
Then if define the length parameter $l$ by
\be
\label{ff2}
l^2={d(d-1) \over \lambda^2}\ ,
\ee
the spacetime metric (\ref{viit}) has the following form:
\be
\label{ff3}
ds_{d+1}^2\sim -{l^2 dy^2 \over y^2} 
+ y\sum_{i,j=1}^d g_{ij}dx^i dx^j\ .
\ee
Thus, the spacetime becomes asymptotically de Sitter one. 
In fact, if we define a time coordinate $t_{\rm asym}$ by 
$y=\e^{2t_{\rm asym} \over l}$, the metric (\ref{ff3}) has 
a warped form:
\be
\label{ff4}
ds_{d+1}^2\sim -d t_{\rm asym}^2 
+ \e^{2t_{\rm asym}}\sum_{i,j=1}^d g_{ij}dx^i dx^j\ .
\ee
Similar solution, which is asymptotically AdS, has been 
found in \cite{NOtwo}. The solution \cite{NOtwo} exists 
when the cosmological constant $\Lambda$ in the action 
(\ref{i}) is negative. The solution in (\ref{xt}) becomes 
asymptotically de Sitter in the time-like infinity but the 
solution in \cite{NOtwo} becomes asymptotically anti-de Sitter 
in the space-like infinity. It would be interesting to construct 
the solution interpolating between asymptotically de Sitter and 
asymptotically Anti-de Sitter ones as in ref.\cite{cvetic}.

We should note that $f(y)$ (\ref{xt}) diverges when 
\be
\label{dS1}
1 - {\alpha c^2 \over \lambda^2 y^d}
 - {kd \over \lambda^2 y}=0\ .
\ee
There are three cases that the above equation has two solutions, 
only one solution and no solution. 
When Eq.(\ref{dS1}) has two solutions ($k\neq 0$), let 
denote the larger solution by $y_0$. When $y\sim y_0$, 
$f(y)$ behaves as 
\be
\label{dS2}
f(y)\sim -{f_0 \over y-y_0}\quad (f_0>0)\ .
\ee
Then if one defines a new (time) coordinate $t$ by 
\be
\label{dS3}
t=2\sqrt{f_0(y - y_0)}\ ,
\ee
the spacetime metric in (\ref{viit}) behaves as 
\be
\label{dS4}
ds_{d+1}^2 \sim - dt^2 + \left({t^2 \over 4f_0} + y_0\right)
\sum_{i,j=1}^d g_{ij}dx^i dx^j\ .
\ee
Then there is obviously no curvature singularity at $y=y_0$ 
($t=0$). Since $t\rightarrow \pm \infty$ corresponds to 
$y\rightarrow +\infty$, the solution connects two asymptotically 
dS regions corresponding to $t\rightarrow \pm \infty$. We also note 
that since the curvature singularity corresponds to $y\rightarrow 0$ 
in (\ref{Axxxviib}), the solution is totally non-singular. 

Let us consider the case that Eq.(\ref{dS1}) has only one solution 
when $k\neq 0$ and $f(y)$ behaves as
\be
\label{dS5}
f(y)\sim -{f_1^2 \over \left(y - y_0\right)^2}\ .
\ee
Then if we define a new (time) coordinate $\tau$ by 
\be
\label{dS6}
\tau = f_1 \ln \left(y-y_0\right)\ ,
\ee
the spacetime metric (\ref{viit}) behaves as 
\be
\label{dS7}
ds_{d+1}^2 \sim - d\tau^2 + \left(\e^{\tau \over f_1} + y_0\right)
\sum_{i,j=1}^d g_{ij}dx^i dx^j\ .
\ee
Then the radius of the universe approaches to a constant 
$\sqrt{y_0}$ when $\tau \rightarrow -\infty$ 
($y\rightarrow y_0$). The curvature singularity corresponding 
to $y\rightarrow 0$ in (\ref{Axxxviib}) does not appear again. 

What happens when $k=0$ and $\alpha>0$? 
For simplicity, we consider $d=4$ case. 
Then Eq.(\ref{dS1}) gives
\be
\label{beta1}
y_0^4={\alpha c^2 \over \lambda^2}\ .
\ee
When $y\sim y_0$, $f(y)$ behaves as (\ref{dS2}). 
If we change the coordinate $y$ to $\theta$ by
\be
\label{beta2}
\cos\theta = {y_0^2 \over y^2}\ ,
\ee
Eq.(\ref{xiiit}) can be integrated explicitly:
\be
\label{beta3}
\phi=\phi_0 + {\theta \over 2}\sqrt{3 \over \alpha}\ .
\ee
Eq.(\ref{beta2}) tells that $\theta=0$ corresponds to $y=y_0$ and 
$\theta\rightarrow {\pi \over 2}$ to $y\rightarrow +\infty$. 
Then we find that $\phi=\phi_0$ at $y=y_0$ and 
$\phi=\phi_\infty\equiv \phi_0 + {\pi \over 4}\sqrt{3 \over \alpha}$. 
Since the string coupling $g_s$ is generally defined by 
\be
\label{beta4}
g_s=\e^{\gamma\phi} \quad (\mbox{$\gamma$ is a constant})\ , 
\ee
the beta function, which may be defined by analogy with AdS/CFT as
\be
\label{beta5}
\beta(\phi) = 2(y-y_0){d g_s \over dy}\ ,
\ee
vanishes at $y=y_0$ and in the limit of $y\rightarrow +\infty$. 
Then $y=y_0$ and $y\rightarrow +\infty$
 correspond to a fixed point of the renormalization group 
from the viewpoint of the dS/CFT correspondence. Then the 
solution of $k=0$ and $\alpha>0$ would connect two fixed points 
where $g_s=\e^{\gamma \phi_0}$ and $g_s=\e^{\gamma \phi_\infty}$. 

The reason why AdS/CFT can be expected is the isometry of 
$d+1$-dimensional anti-de Sitter space, which is $SO(d,2)$ 
symmetry. It is identical with the conformal symmetry of 
$d$-dimensional Minkowski space. We should note, however,  
$d+1$-dimensional de Sitter space has the isometry of 
$SO(d+1,1)$ symmetry, which can be a conformal symmetry of 
$d$-dimensional Euclidean space. Then it might be natural to 
expect the correspondence between $d+1$-dimensional de Sitter 
space and $d$-dimensional euclidean conformal symmetry (dS/CFT 
correspondence\cite{strominger}). 
In fact, the metric of 
$D=d+1$-dimensional anti de Sitter space (AdS) is given by
\be
\label{AdSm}
ds_{\rm AdS}^2=dr^2 + \e^{2r}\left(-dt^2 + \sum_{i=1}^{d-1}
\left(dx^i\right)^2\right)\ .
\ee
In the above expression, the boundary of AdS lies at 
$r=\infty$. If one exchanges the radial coordinate $r$ 
and the time coordinate $t$, we obtain the metric of the 
de Sitter space (dS): 
\be
\label{dSm}
ds_{\rm dS}^2=-dt^2 + \e^{2t}\sum_{i=1}^d
\left(dx^i\right)^2\ .
\ee
Here $x^d=r$. Then there is a boundary at $t=\infty$, where the 
Euclidean conformal field theory (CFT) can live and one expects 
dS/CFT correspondence as one more manifestation of holographic 
principle.
Strominger \cite{strominger} has conjectured that dual CFT in dS/CFT 
may be non-unitary. 

Using the above indication to dS/CFT correspondence, one can speculate
on the properties of our asymptotically dS solutions in terms of RG flow.
Indeed, the above analysis seems to tell that there is a renormalization 
group 
flow. In case of the AdS$_5$/CFT$_4$ correspondence, the curvature radius 
$l_{\rm AdS}$, by which the scalar curvature is given by 
$R=-{20 \over l_{\rm AdS}^2}$, can be expressed by the string 
coupling and the integer $N$ which parametrizes the gauge 
group $SU(N)$ or $U(N)$ of the CFT as follows
\be
\label{c1}
l=\left(4\pi g_s N\right)^{1 \over 4}\ .
\ee
Then it might be one of natural ways to define the $c$-function, 
in the situation under consideration in this paper, by the scalar curvature:
\be
\label{c2}
c=\left(R\right)^{-{3 \over 2}}\ .
\ee
Since Eq.(\ref{iit}) gives, for $d=4$, 
\be
\label{c3}
R={5 \over 3}\Lambda + \alpha \left(\partial_y\phi\right)^2\ ,
\ee
by using the solution (\ref{xt}) with $d=4$ and $k=0$, one has 
\be
\label{c4}
c=\left({5 \over 3}\lambda^2 + {3y_0^4 \over y^6
\left(1 - {y_0^4 \over y^4}\right)}\right)^{-{3 \over 2}}\ .
\ee
Here $y_0$ is defined by (\ref{beta1}).
Then $c$ is the monotonically increasing function with respect 
to $y$ and it vanishes when $y=y_0$. From the viewpoint of the dual 
field theory, the vanishing $c$ means that all the fields 
become massive. In the AdS/CFT correspondence, the dilaton gravity 
with the constant potential, which can be identified with the 
cosmological constant, as in (\ref{i}), corresponds to the 
deformation of the conformal field theory by the marginal 
operator but the above non-trivial flow would tell that the 
deformation is not exactly marginal. 

We can extend the definition of the $c$-function 
(\ref{c2}) to the case that there is a non-trivial potential 
$V(\phi)$ instead of the cosmological term, that is, $\Lambda$ 
in (\ref{i}) is replaced by $V(\phi)$. If we assume the metric 
has the form of (\ref{viit}) and $\phi$ only depends on $y$, we
have 
\be
\label{c5}
R={5 \over 3}V(\phi) + \alpha \left(\partial_y\phi\right)^2\ ,
\ee
instead of (\ref{c3}) or (\ref{iit}) and the equation 
corresponding to (\ref{iiit}) has the form 
\be
\label{c6}
2\alpha\partial_y^2\phi = V'(\phi)\ .
\ee
By using (\ref{c5}) and (\ref{c6}), one finds
\be
\label{c7}
\partial_y R = \left({d+1 \over d-1}V'(\phi) + 2\alpha
\partial_y^2\phi\right)\partial_y\phi
={2 \over d-1}\partial_y\left(V(\phi)\right)
={2\alpha \over d-1}\partial_y\left(\left(\partial_y
\phi\right)^2\right)\ .
\ee
Then the $c$-function defined by (\ref{c2}) becomes stationary 
when $V'(\phi)=0$. 
We also note that  when $\partial_y\phi=0$, the $c$-function 
defined by (\ref{c2}) has the following form  
\be
\label{c8}
c\propto \left(V(\phi)\right)^{-{3 \over 2}}\ ,
\ee
which is similar to the standard form $c_{\rm AdS}= 
\left(-V(\phi)\right)^{-{3 \over 2}}$ in the AdS/CFT 
correspondence \cite{GPPZ,BGM}. 
If we choose the metric for $k=0$ and $d=4$ in the following form, 
instead of (\ref{viit}),  
\be
\label{c9}
ds_{d+1}^2= - dt^2 + \e^{2A(t)}\sum_{i=1}^4 dx_i dx^i\, 
\ee
the Einstein equations are:
\bea
\label{c10}
{d^2 A \over dt^2}&=& - {\alpha \over 3}\left({d\phi \over dt}
\right)^2 \ ,\\
\label{c11}
\left({dA \over dt}\right)^2 &=& {V(\phi) \over 12} 
+ {\alpha \over 12}\left({d\phi \over dt}
\right)^2 \ .
\eea
Eq.(\ref{c10}) tells that ${dA \over dt}$ is a monotonically 
decreasing function of $t$. We also note that the region where 
${dA \over dt}\sim 0$ is asymptotically de Sitter space. It 
is natural, as in the case of  AdS/CFT correspondence 
\cite{FGPW}, to assume that  $c$-function is given by 
\be
\label{c12}
c \propto \left({dA \over dt}\right)^{-{3 \over 2}}\ .
\ee
Eq.(\ref{c10}) indicates that at the critical point, where 
${dc \over dt}=0$ $\left({d^2 A \over dt^2}=0\right)$, we 
have ${d\phi \over dt}=0$, as is assumed in (\ref{c8}). Then 
Eq.(\ref{c11}) tells that 
\be
\label{c13}
\left({dA \over dt}\right)^2 = {V(\phi) \over 12} 
\ee
at the critical point, what proves that Eq.(\ref{c8}) 
and therefore the definition of the $c$-function here 
(\ref{c2}) is consistent. Of course, in the absence of the explicit
example of dual CFT all above discussion on $c$-function is purely 
speculation. 
Nevertheless, the existence of consistent $c$-function for dual RG flow 
gives further support to dS/CFT correspondence.

In \cite{BST}, it has been shown that there is an S$^{8-p}$ 
compactification of the $D=10$ supergravity theory to an 
effective gauged $(p+2)$-dimensional supergravity with the action
\bea
\label{dS8}
S&=&N^2 \int d^{p+2}x \sqrt{-g} \left(N\e^\phi\right)^{2(p-3)
\over (7-p)} 
\left[R + {4(p-1)(p-4) \over (7-p)^2}\partial_\mu\phi 
\partial^\mu \phi \right.\nn
&& \left. + {1 \over 2}(9-p)(7-p)\right]\ . 
\eea
Here $N$ expresses the flux of the $(8-p)$-form, which 
generates the effective cosmological constant. 
In the Einstein frame, the action (\ref{dS8}) can be rewritten 
in the following form:
\bea
\label{dS9}
S&=&N^2 \int d^{p+2}x \sqrt{-g} 
\left[R - {1 \over 2}\partial_\mu\phi 
\partial^\mu \phi \right.\nn
&& \left. + {1 \over 2}(9-p)(7-p)N^{4(p-3) \over p(p-7)} 
\e^{-{\sqrt{2} (p-3) \over \sqrt{p(9-p)}}\phi}\right]\ .
\eea
Then $p=3$ case is similar to the action (\ref{i}) with 
$d=4$ except the sign of the cosmological term. 
It has been shown by Hull \cite{ds1} that if we consider Type 
IIB$^*$ string theory, which is given by the time-like T-duality 
from Type II string theory, D-brane is transformed into E-brane 
and its near light-cone limit gives the de Sitter space instead of 
the AdS space, which is the near horizon limit of D-brane. 
In the time-like T-duality, the sign in front of the kinetic 
term of $(8-p)$-form is changed. Then in the effective theory as 
in (\ref{dS8}) or (\ref{dS9}), the sign of the cosmological 
constant is changed, what corresponds to the action 
(\ref{i}). 

As the solution (\ref{xt}) is asymptotically de Sitter space, 
one can consider the renormalization flow as in \cite{NOds} in 
an analogy with the holographic renormalization group 
formulation for AdS/CFT developed in
\cite{BVV,FMS}. Take $D=d+1$ dimensional dS-like 
metric in the following form 
\be
\label{met}
ds^{2} = G_{\mu\nu}dX^\mu dX^\nu = 
-dt^{2} + \hat g_{ij}(x,t)dx^idx^j .
\ee
where $X^\mu=(x^i,t)$ with $i,j=1,2,\cdots ,d$. 
We consider the action which is the sum of the action $S$  
(\ref{i}) on a $(d+1)$ dimensional manifold $M_{d+1}$ and 
the action on the boundary $\Sigma_{d}=\partial M_{d+1}$, 
\bea
\label{ib}
S_{d+1}&=& S + 
2\int_{\Sigma_{d}} d^{d}x \sqrt{\hat g}K \nn
&\equiv & \int d^{d}x dt \sqrt{-G} {\cal L}_{d+1}.
\eea
where $K_{\mu\nu}$ is the extrinsic curvature on $\Sigma_{d}$.
The action is taken in the 
Minkowski signature. 
Since we are considering the de Sitter background instead of 
the AdS background, the cosmological constant $\Lambda$ is 
positive and parametrized by the parameter $l$, which is the 
radius of the asymptotic dS$_{d+1}$
\be
\label{dSL}
\Lambda={d(d-1) \over l^2}\ .
\ee
In the metric (\ref{met}), $K_{\mu\nu}$ is given as
\bea
\label{iib}
K_{ij}={1\over 2}\partial_t \hat g_{ij}\ , \quad
K= \hat g^{ij}K_{ij}
\eea
In the canonical formalism, ${\cal L}_{d+1}$ is rewritten
by using the canonical momenta $\Pi_{ij}$ and $\Pi$ 
and Hamiltonian density ${\cal H}$ as
\bea
\label{iiib}
{\cal L}_{d+1} &=& \Pi^{ij}\partial_t \hat g_{ij} 
+ \Pi\partial_t\phi +{\cal H} \ ,\nn
{\cal H} &\equiv& {1 \over d-1}(\Pi ^{i}_{\ i})^2
 -\Pi_{ij}\Pi^{ij} -{1 \over 2\alpha} \Pi^2 + \hat R-\Lambda 
 - \alpha \hat g^{ij}\partial_i \phi \partial_j \phi \ . 
\eea
The equation of motion for $\Pi^{\mu\nu}$ leads to 
\be
\label{iv}
\Pi^{ij}=K^{ij}-\hat g^{ij}K\ ,\quad 
\Pi=\alpha \partial_t\phi\ .
\ee
The Hamilton constraint ${\cal H}=0$ leads to the
Hamilton-Jacobi equation (flow equation) 
\bea
\label{HJ}
\{ S,S \} (x) &=& \sqrt{\hat g} {\cal L}_{d} (x) \\
\{ S,S \} (x) &\equiv & {1 \over \sqrt{\hat g}}
\left[-{1 \over d-1}\left(\hat g_{ij}{\delta S 
\over \delta \hat g_{ij}}  \right)^{2}
+ {\delta S \over \delta \hat g_{ij}}
{\delta S \over \delta \hat g^{ij}} 
+ \alpha \left({\delta S \over \delta \phi}\right)^2 \right] , \\
{\cal L}_{d}(x) &\equiv & \hat R[\hat g] - \Lambda 
-\alpha \hat g^{ij}\partial_i \phi \partial_j \phi\ .
\eea
One can decompose the action $S$ into a local and non-local part as
discussed in ref.\cite{BVV} as follows
\bea
\label{vb}
S[\hat g(x)] &=& S_{loc}[\hat g(x)]+\Gamma[\hat g(x)] ,
\eea
Here $S_{loc}[\hat g(x)]$ is tree level action and $\Gamma$ contains
the higher-derivative and non-local terms. 
In the following discussion, we take the systematic method 
of ref.\cite{FMS}, which is weight calculation. 
The $S_{loc}[\hat g]$ can be expressed as a sum of local terms
\be
\label{via}
S_{loc}[\hat g(x)] = \int d^{d}x \sqrt{\hat g} {\cal L}_{loc}(x) 
= \int d^{d}x \sqrt{\hat g} 
\sum _{w=0,2,4,\cdots} [{\cal L}_{loc}(x)]_{w} 
\ee
The weight $w$ is defined by following rules;
\be
\label{weight}
\hat g_{ij},\; \Gamma : \mbox{weight 0} \ ,\quad
\partial_{\mu} : \mbox{weight 1} \ ,\quad
\hat R,\; \hat R_{ij} : \mbox{weight 2} \ ,\quad
{\delta \Gamma \over \delta \hat g_{ij}} : 
\mbox{weight $d$} \ .
\ee
Using these rules and (\ref{HJ}), one obtains 
the equations, which depend on the weight as
\bea
\label{wt1}
\sqrt{\hat g}{\cal L}_{d} 
&=& \left[ \{ S_{loc},S_{loc} \} \right]_0 +
\left[ \{ S_{loc},S_{loc} \} \right]_2 \\
\label{wt2}
0 &=& \left[ \{ S_{loc},S_{loc} \} \right]_w  
\quad (w=4,6,\cdots d-2), \\
\label{wt3}
0 &=& 2\left[ \{ S_{loc}, \Gamma \} \right]_d 
+ \left[ \{ S_{loc},S_{loc} \} \right]_d
\eea
The above equations which determine 
$\left[ {\cal L}_{loc} \right]_{w}$. 
$\left[ {\cal L}_{loc} \right]_{0}$ and $[{\cal L}_{loc}]_{2}$ are
parametrized by 
\be
\label{vi}
\left[ {\cal L} _{loc} \right]_0 = W\ ,\quad 
\left[ {\cal L} _{loc} \right]_2 = -\Phi \hat R 
+ M \hat g^{ij}\partial_i \phi\partial_j \ .
\ee
Thus one can solve (\ref{wt1}) as
\bea
\label{vii}
\Lambda = {d \over 4(d-1)}W^{2} \ ,\quad 
1 = {d-2 \over 2(d-1)} W\Phi\ \quad
\alpha=-{d-2 \over 4(d-1)}WM
\ .
\eea
The case of $d=2$ is special and instead of 
Eqs.(\ref{wt1},\ref{wt2},\ref{wt3}), we obtain
\be
\label{d2ii}
\sqrt{\hat g}{\cal L}_2 = 
2\left[ \{ S_{loc}, \Gamma \} \right]_2 
+ \left[ \{ S_{loc},S_{loc} \} \right]_0 +
\left[ \{ S_{loc},S_{loc} \} \right]_2 \ .
\ee
When $d=2$, the second equation  
in (\ref{vii}) is irrelevant but by using (\ref{dSL}), 
we obtain
\be
\label{d2i}
W_2=- {2 \over l}\ . 
\ee
When $d>2$, by using (\ref{dSL}), one obtains $W$ and $\Phi$ as
\be
\label{viii}
W=-{2(d-1) \over l}\ ,\quad \Phi = -{l \over d-2}\ ,
\quad M = {2l\alpha \over d-2}\ .
\ee
Note that there is an ambiguity in the choice of the 
sign but the relative sign of $W$ and $\Phi$ is different 
from AdS case. 

When $\left[ {\cal L}_{loc} \right]_{4}=0$, one gets 
for $d=4$ and $2\alpha=1$
\bea
\label{ano4}
\lefteqn{{1 \over \sqrt{\hat g}}\left[\{ S_{loc},S_{loc} \} 
\right]_{4}} \nn
&&={l^2 \over 12}
\hat R^{2}  - {l^2 \over 4}\hat R_{ij}\hat R^{ij} 
 - {l^2 \over 12} \hat R \hat g^{ij}\partial_i\phi\partial_j\phi 
 + {1 \over 4}\hat R^{ij}\partial_i\phi \partial_j\phi \nn
&& - {1 \over 24}\left(\hat g^{ij}\partial_i\phi\partial_j\phi\right)^2 
 - {1 \over 8}\left\{{1 \over \sqrt{\hat g}}\partial_i\left(
 \sqrt{\hat g} \hat g^{ij}\partial_j\phi\right)\right\}^2  \ . 
\eea
Then 4d holographic conformal anomaly is obtained as follows:
\bea
\label{xi}
\kappa^2{\cal W}_{4} &=& {l\over 2\sqrt{\hat g} }
\left[ \{S_{loc},S_{loc} \}\right]_{4} \nn
&=& l^{3}\left[ {1\over 24}\hat R^{2} -
{1\over 8} \hat R_{ij}\hat R^{ij}
 - {l^2 \over 24} \hat R \hat g^{ij}\partial_i\phi\partial_j\phi 
 + {1 \over 8}\hat R^{ij}\partial_i\phi \partial_j\phi \right.\nn
&& \left. 
 - {1 \over 48}\left(\hat g^{ij}\partial_i\phi
 \partial_j\phi\right)^2 
 - {1 \over 16}\left\{{1 \over \sqrt{\hat g}}\partial_i
 \left(\sqrt{\hat g} \hat g^{ij}\partial_j\phi\right)
 \right\}^2 \right]\ .
\eea
The obtained form of the anomaly is  identical with 
that in \cite{anomNO}. 
Thus as an extension of ref. \cite{NOds} in the dS/CFT 
correspondence, we reproduced the dilatonic conformal anomaly, 
which was obtained from the AdS/CFT framework in \cite{anomNO}.
(There exists extensive list of refs. related with the study of
holographic anomaly in AdS/CFT \cite{anom}).
The Weyl anomaly coming from the multiplets of ${\cal N}=4$ 
supersymmetric $U(N)$ or $SU(N)$ Yang-Mills coupled with ${\cal N}=4$ 
conformal supergravity was calculated in \cite{HT}. 
The expression in \cite{anomNO} as well as the one  
(\ref{xi}) corresponds to the anomaly of the supergravity in 
the background where only gravity and the real part of the 
scalar field $\varphi$ from the ${\cal N}=4$ conformal supergravity 
multiplet are non-trivial and other fields vanish. Thus, holographic RG 
flow equations permit to calculate 4d conformal anomaly from 5d
asymptotically  dS space. This maybe considered as further support of 
proposed dS/CFT correspondence.

Finally, let us make several remarks on possibility of dS supergravities.
The dS/CFT can be conjectured from the Type II$^*$ string 
\cite{ds1}, which is obtained from Type II string theory by the 
time-like $T$-duality. 
A solution of Type II supergravity is given by
\bea
\label{Hi}
ds^2 &=& H^{-{1 \over 2}}\left(-dt^2 + dx_1^2 + ... + dx_p^2\right)
+ H^{1 \over 2}\left(dx_{p+1}^2 + ... + dx_9^2\right) \nn
&& H=c + {q \over r^{7-p}}\ ,\quad r^2\equiv 
\sum_{i=p+1}^9 x_i^2\ .
\eea
If we consider near horizon limit ($r^2\rightarrow 0$) in case 
$p=3$, which corresponds to D3-brane, the metric 
(\ref{Hi}) has the following form:
\be
\label{Hvi}
ds^2 = {r^2 \over a^2}\left(-dt^2 + dx_1^2 + ... + dx_3^2\right)
+ {a^2 \over r^2}dr^2 + a^2 d\Omega_5^2\ .
\ee
Here $d\Omega_5^2$ is the metric of the 5-dimensional unit 
sphere. Then the 10 dimensional spacetime becomes 
AdS$_5\times $S$_5$. Taking the time-like $T$-duality, 
the component $g_{tt}$ of the metric tensor is replaced by 
${1 \over g_{tt}}$. Then from (\ref{Hi}), one can obtain a 
solution in Type II$^*$ supergravity:
\be
\label{Hv}
ds^2 = H^{-{1 \over 2}}\left(dx_1^2 + ... + dx_p^2\right)
+ H^{1 \over 2}\left(-dt^2 + dx_{p+1}^2 + ... + dx_9^2\right)
\ee
Considering near light-cone limit 
($\tau^2\rightarrow 0$), instead of the near horizon limit, and 
changing the coordinate by $t=\tau\cosh \beta$ and 
$r=\tau\sinh\beta$,  assuming $\tau^2=t^2 - r^2>0$, we obtain 
the following metric 
\bea
\label{Hviii}
&& ds^2 = {\tau^2 \over a^2}\left(-dt^2 + dx_1^2 + ... 
+ dx_3^2\right) - {a^2 \over \tau^2}d\tau^2 
+ a^2 d\tilde\Omega_5^2\ , \nn
&& d\tilde\Omega_5^2=d\beta^2 + \sinh^2\beta d\Omega_4^2\ .
\eea
Here $d\tilde\Omega_5^2$ is the metric of 5-dimensional 
hyperboloid. Then the spacetime in (\ref{Hviii}) is the 
product of the hyperboloid and the 5-dimensional 
de Sitter space, whose metric is ${\tau^2 \over a^2}
\left(-dt^2 + dx_1^2 + ... + dx_3^2\right) - {a^2 \over 
\tau^2}d\tau^2$. Even in case $\tau^2=-\sigma^2=t^2 - r^2<0$, 
by the similar calculation, we obtain the spacetime which is 
the product of the hyperboloid and the 5-dimensional 
de Sitter space.
The expression (\ref{xi}), therefore, might 
indicate that supergravity obtained from Type II$^*$ string 
has the common structure with the supergravity obtained from 
Type II string.  

\ 

\noindent
{\bf Acknoweledgements} 
The work by SN is supported in part by the Ministry of Education, 
Science, Sports and Culture of Japan under the grant n. 13135208.
The research by SDO is supported in part by RFBR and in part by
GCFS Grant E00-3.3-461.

\end{document}